\documentclass[letterpaper,english,aps, nofootinbib, prd, twocolumn, lengthcheck, superscriptaddress]{revtex4-2}
\usepackage[T1]{fontenc}

\usepackage[colorlinks=true,linktocpage=true,linkcolor=blue,citecolor=blue]{hyperref}
\usepackage{verbatim}
\usepackage{bm}
\usepackage{amsmath}
\usepackage{amssymb}
\usepackage{graphicx}
\usepackage{slashed}
\usepackage{wrapfig}
\usepackage{bbm}
\usepackage[caption=false]{subfig}
\usepackage{verbatim}
\newcommand \beq{\begin{eqnarray}}
\newcommand \eeq{\end{eqnarray}}

\newcommand{\rmd}{\mathrm{d}}

\begin{document}
\title{QCD equations of state and speed of sound in neutron stars}
\author{Toru Kojo}
\affiliation{Key Laboratory of Quark and Lepton Physics (MOE) and Institute of Particle Physics, Central China Normal University, Wuhan 430079, China}  
\date{\today}

\begin{abstract}
Neutron stars are cosmic laboratories to study dense matter in Quantum Chromodynamics (QCD). 
The observable mass-radius relations of neutron stars are determined by QCD equations of state, and can reflect the properties of QCD phase transitions.
In the last decade there have been historical discoveries in neutron stars, the discoveries of two-solar mass neutron stars and neutron star merger events, which have imposed tight constraints on equations of state. 
While a number of equations of state are constructed to satisfy these constraints, a theoretical challenge is how to reconcile those constructions with the microphysics expected from the hadron physics and in-medium calculations.
In this short article we briefly go over recent observations and discuss their implications for dense QCD matter, referring to QCD constraints in the low and high density limits, QCD-like theories, and lattice QCD results for baryon-baryon interactions.

\end{abstract}

\maketitle

\section{Introduction}
\label{sec:intro}

Quantum Chromodynamics (QCD) is the theory of strong interactions with quarks and gluons being elementary degrees of freedom (d.o.f.) \cite{Gross:1973id,Politzer:1973fx}. In a normal circumstance they are confined into hadrons, e.g., protons, neutrons, pions, and so on. These composite particles are effective d.o.f. to describe the physics of nuclei or the vacuum. The situation changes when we heat a gas of hadrons or compress matter made of nuclei; when hadrons overlap, quarks and gluons become natural d.o.f. to describe the physics in extreme environments \cite{Fukushima:2010bq}. These extreme matters appear in the early universe and in astrophysical objects such as neutron stars \cite{yagi2005quark}.

Our understandings seem matured for some extreme environments. In hot QCD, a heated hadron resonance gas (HRG) continuously transforms into a quark-gluon-plasma (QGP), in spite of apparent differences of these two phases. Experimentally such hot matter has been created by heavy ion collisions at high energy in which the colliding energy is converted into heat. Combined analyses of the experiments, ab-initio lattice simulations of QCD, and model calculations, together have formed a plausible picture of hot QCD matter \cite{Andronic:2017pug}.

Another extreme environment is cold, dense matter of QCD (for reviews, see, e.g. \cite{Baym:2017whm,Buballa:2014tba,Alford:2007xm}). Compressing matter but keeping temperature low should lead to quark matter, as  proposed in 70's \cite{Itoh:1970uw,Collins:1974ky}. How nuclear matter changes into quark matter is still unknown, as theoretical framework for such transitions is not established; in particular the sign problem prevents us from performing the lattice Monte-Carlo simulations. Only exceptions are (i) the domain of nuclei around baryon density, $n_B =n_0 \simeq 0.16\,{\rm fm}^{-3}$, where we have nuclear experiments, and (ii) very high density, $n_B \sim 40\, n_0$, where the asymptotic freedom in QCD allows weak coupling calculations \cite{Freedman:1976ub,Kurkela:2009gj}. Cold matter denser than nuclear matter cannot be created in terrestrial experiments such as heavy ion collisions which inevitably produce heat.

In this respect neutron stars (NSs) are unique cosmic laboratories to study cold, dense matter of QCD beyond the nuclear regime. A NS is an extremely dense object which contains a solar mass ($M_\odot$) within the radius of $\sim 12$ km  \cite{glendenning2012compact}. Having a very large energy within a small domain, the gravitational is very strong. A star would collapse into a blackhole (BH) unless the gravitational force is balanced with the matter pressure. The structure of NSs is determined by the Einstein equation coupled to the QCD (plus electroweak) energy-momentum tensor, which is reduced to the Tolman-Oppenheimer-Volkoff (TOV) equation for a spherical star. The electroweak interactions affect the matter composition through the charge neutrality and $\beta$-equilibrium conditions on equations of state (EoS). The mass-radius ($M$-$R$) relations of NSs are the most basic observables which have one-to-one correspondence with neutron star EoS where QCD plays dominant roles. 

The observational determination of $M$-$R$ curves is not straightforward, as we have to look for good signals from the universe. Nevertheless, the last decade had  historical discoveries: the existence of two-solar mass ($2M_\odot$) pulsars \cite{Demorest:2010bx,Fonseca:2016tux,Arzoumanian:2017puf,Antoniadis:2013pzd,Cromartie:2019kug} and a detection of gravitational waves (GWs) from a NS merger (GW170817) \cite{TheLIGOScientific:2017qsa} with the associated electromagnetic (EM) counterparts \cite{Goldstein:2017mmi,Savchenko:2017ffs,Monitor:2017mdv,GBM:2017lvd,Coulter:2017wya,Troja:2017nqp,Hallinan:2017woc}. These findings allow us to delineate the properties of dense QCD matter. The on-going and up-coming observational programs will be listed in Sec.\ref{sec:obs}.

We are getting better constraints on $M$-$R$ relations or EoS, but they do not directly tell the matter composition. We need to look into the microphysics behind EoS, taking into account the constraints from the nuclear physics near the saturation density and the finite size effects of hadrons. Taking typical estimates on these quantities (reviewed in Sec.\ref{sec:eos}), we can infer the density domain for a given point of $M$ and $R$. For a canonical NS with the mass $\sim 1.4 M_\odot$, the density is $2$-$3n_0$, and hence may be characterized within the hadronic regime, while for $2M_\odot$ NSs the core density may go beyond the density, $n_B \sim 5n_0$, at which baryons of the radii 0.5-0.8 fm overlap, and hence the core may accommodate quark matter. If a massive NS indeed contains quark matter, there should be hadron-to-quark matter phase transitions near the core. The nature of the phase transition has not been determined from the current observations; modeling based on different orders of phase transitions can be arranged to pass the observational constraints (see Sec. \ref{sec:cs2}). But the observations have certainly constrained the strength of the phase transition and our model building. The first order phase transition, if exists at low temperatures, would imply the existence of the QCD critical end point phase at finite temperature; this is one of important targets in the beam energy scan program of heavy ion experiments  \cite{Fukushima:2020yzx,Dexheimer:2020zzs}.

In this article we begin with reviewing observational constraints (Sec.\ref{sec:obs}), and combine them with theoretical considerations on matter properties (Sec.\ref{sec:eos}). Particular attention is payed for the speed of sound which differentiates dense QCD matter from conventional matters (Sec.\ref{sec:cs2}). We take the natural unit $c=\hbar=k_B=1$.

\section{Observational constraints}
\label{sec:obs}

\begin{figure}[tb]
\vspace{0.0cm}
\centering
\includegraphics[scale=0.21]{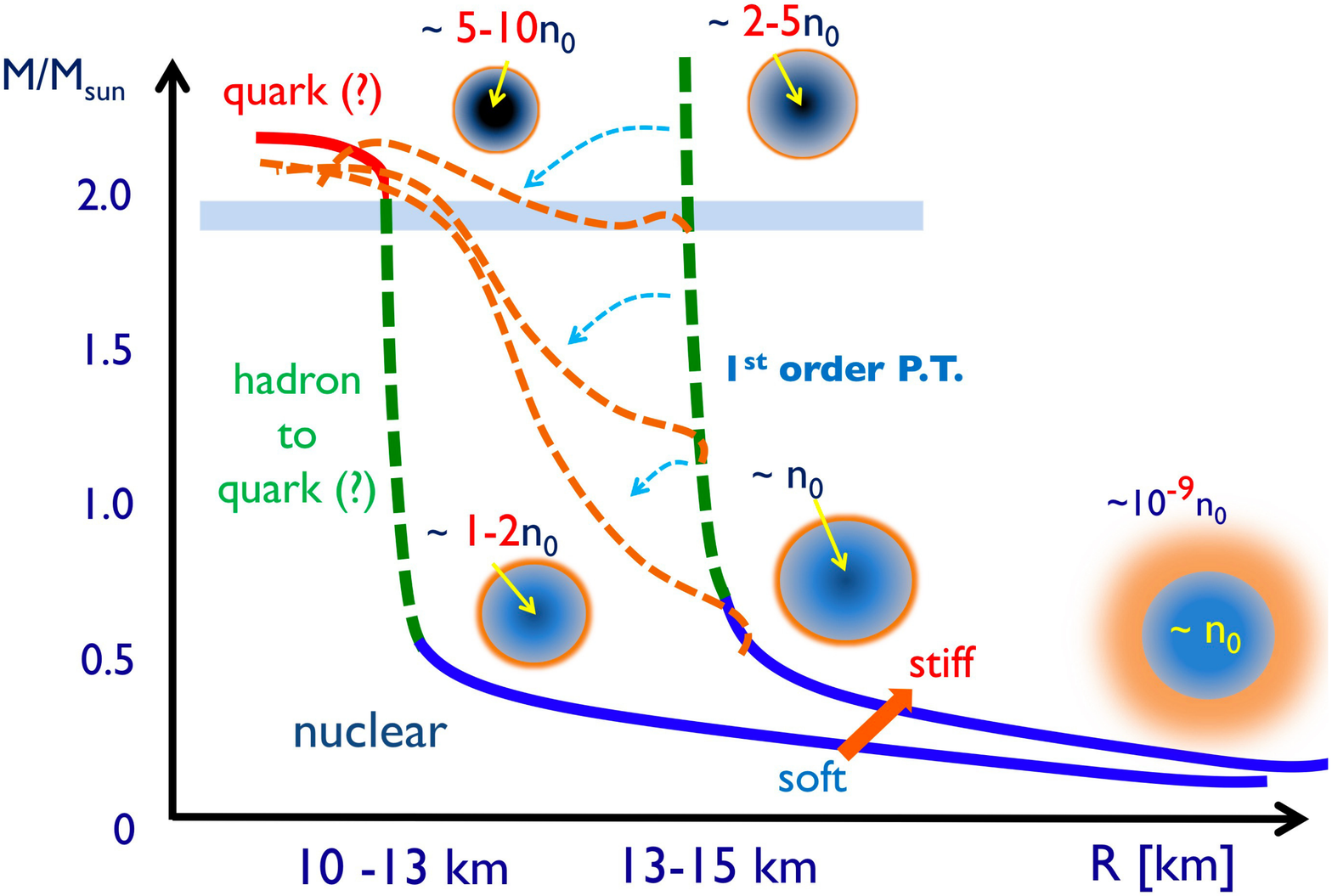}
\caption{
The correlation between the $M$-$R$ relation and equations of state.
  }
  \vspace{-0.5cm}
   \label{fig:M-R}
\end{figure}

{\it $M$-$R$ relations ---} In order to obtain a $M$-$R$ point, we assign a core density $n_c$, and then integrate matter from the core to the surface until the pressure reaches zero, $P(r=R)=0$ \cite{glendenning2012compact}. The $M$-$R$ points at different core densities form a $M$-$R$ curve for a given EoS. $M$-$R$ curves also depend on rotations, but in most cases they can be treated in perturbative treatments \cite{Hartle:1967he,Hartle:1968si}. But the rotation goes beyond the perturbative regime for objects after the NS mergers (see below). For later convenience we call the maximum mass of a non-rotating star $M_{\rm TOV}$, and the radius of $1.4 M_\odot$ NS $R_{1.4}$.

In NSs, EoS at various densities contribute. But the shape of $M$-$R$ curves can be largely characterized by EoS at fiducial densities \cite{Lattimer:2000nx}. Shown in Fig. \ref{fig:M-R} is a typical $M$-$R$ curve. A low mass star has a core plus a loosely bound crust. The rapid reduction of $R$ within a small increase in $M$ is due to the compression of dilute crust matter. For heavier stars the crust is very thin, and the size of the dense core, which is hard to compress, is observed. This turning point is seen around $n_c \sim n_0$. The curve passes the domain of canonical $1.4M_\odot$ NSs at $n_c \sim 2$-$3n_0$, and reaches $M > 2M_\odot$ at $n_c \gtrsim 5n_0$.

This illustration suggests that the overall radii of NSs are determined by EoS at low density, $n_B = 1$-$3n_0$, while $M_{\rm TOV}$ by EoS at high density, $n_B\gtrsim 5n_0$. Now we list up observational constraints.
\\
\\
{\it $2M_\odot$ Pulsars ---} One of the historical measurements in the last decade was the Shapiro delay measurement of the pulsar PSR J1614-2230, announced in 2010. The mass was initially estimated as $1.97^{+ 0.04}_{-0.04} M_\odot $ \cite{Demorest:2010bx}, and after long term observations the estimate has been updated to $1.908^{+ 0.016}_{-0.016} M_\odot $ (68\%CL) \cite{Arzoumanian:2017puf}. The second precisely measured $2M_\odot$ NS is PSR J0348+0432 whose mass is $2.01^{+0.04}_{-0.04} M_\odot$ \cite{Antoniadis:2013pzd}. A new Shapiro delay measurement was done for the PSR J0740+6620 where the mass is estimated to be $M=2.14^{+0.10}_{-0.09} M_\odot$ \cite{Cromartie:2019kug}. These results have established the $2M_\odot$ constraints that require high density EoS to be stiff.
\\
\\
{\it NS mergers ---} Another historical event is a NS merger, GW170817, from which GWs were detected by the aLIGO and Virgo \cite{TheLIGOScientific:2017qsa}. Furthermore, this event happened at a rather close distance, $43.8^{+2.9}_{-6.9}$ Mpc, allowing the measurements of  EM signals \cite{Goldstein:2017mmi,Savchenko:2017ffs,Monitor:2017mdv,GBM:2017lvd,Coulter:2017wya,Troja:2017nqp,Hallinan:2017woc}, with which the merger event was analyzed from various aspects. The observations have been compared to general relativistic numerical simulations for various EoS \cite{Kiuchi:2019kzt,Narikawa:2019xng}. 

The merger event experiences various stages, and each of which has distinct signals (for a review, e.g. Ref. \cite{Radice:2020ddv}). In an {\it early spiral} phase two NSs are widely separated, and each can be treated as a point particle. This stage informs us the total mass. With long term emissions of GWs, two NSs come close enough to deform the shape of NSs. This is a {\it tidally deformed} phase. The deformation of the NSs adds an extra gravity whose net effect is attractive and hence accelerates the merging process. This affects the GW form patterns. The degree of the deformation is characterized by the tidal deformability which has strong correlations with a NS radius; a NS with a larger radius is more easily deformed, leading to a larger tidal deformability. In the end NSs collide. The stage after the collision is called a {\it post merger} phase, a highly dynamical stage. There are various signals; the gamma-ray burst, blue- and red-ejecta, neutrinos, and GWs at high frequency, 1-4 kHz, which reflects rapid oscillations of a compact object \cite{Hotokezaka:2011dh,Takami:2014zpa,Weih:2019xvw}. At beginning, the merged object is differentially rotating, and this rotation prevents the object from immediate gravitational collapse. After short time this differential rotation is braked by viscous effects and magnetic field effects. After this loss of centrifugal effects, the object exceeding the maximum NS mass collapses to a BH. The ejecta in a post-merger phase is sensitive to the compactness of merged objects and the time scale of the gravitational collapse; for example the amount of ejecta would be too little if the merger immediately collapses or is too compact.

In the event GW170817, the total mass before the merger was estimated in good accuracy, 2.73-2.78$M_\odot$, and the GWs in early spiral and tidally deformed phases were measured \cite{TheLIGOScientific:2017qsa}. Referring to GW templates from numerical simulations, the tidal deformability (or radii of neutron stars) was constrained. The aLIGO-Virgo collaboration estimated radius to be $11.9^{+1.4}_{-1.4}$ km for each star \cite{Abbott:2018exr}, which is largely consistent with estimates by other studies \cite{Annala:2017llu,De:2018uhw}. The current detector did not have enough sensitivity to detect the GWs in the post merger phase. But the EM detection provides information of the post merger phase; the estimated amount of ejecta of $\sim 10^{-2}M_\odot$, which is correlated with the lifetime and compactness of the merger,  is used to put the upper- and lower-bound on the $M_{\rm max}$, and the lower-bound on $R_{1.4}$ \cite{Radice:2017lry}, although the bounds depend on our interpretation on the fate of the merger. The first and standard one is the hypermassive NS (HMNS) scenario, in which the merger becomes a metastable star with differential rotations and collapses. 
The second one is the supramassive NS (SMNS) scenario, in which a HMNS changes into a long-lived, rigidly rotating NS whose maximal mass is greater than the corresponding non-rotating star by a factor $1.2$. The threshold mass $ M_{\rm th} = (2.73$-$2.78M_{\odot}$)$/1.2 \simeq 2.28$-$2.32 M_\odot$ appears as $M_{\rm TOV} < M_{\rm th}$ in the HMNS scenario \cite{Margalit:2017dij,Ruiz:2017due,Rezzolla:2017aly,Shibata:2019ctb}, while as $M_{\rm TOV} > M_{\rm th}$ in the SMNS scenario \cite{Yu:2017syg,Piro:2018bpl}. Based on the HMNS picture, the lower-bound of the radius was also estimated \cite{Bauswein_2017,Radice:2017lry}, although later it was found that the condition can be significantly relaxed by allowing a sufficiently large $M_{\rm TOV}$ for the long life time \cite{Kiuchi:2019lls}.

After the discovery of GW170817, several (candidates of) NS mergers have been found. The second detection is the GW190425 event with a large total mass $3.4^{+0.3}_{-0.1} M_\odot$ \cite{Abbott:2020uma}. This system is exotic because at least one of NSs should be different from canonical NSs with $1.4M_\odot$. The possibility of BH-NS binary is also not excluded \cite{Kyutoku:2020xka,Han:2020qmn}. The EM counterparts have not been detected and the tidal deformability is not well constrained. This is in part because this event happens at a far distance, $159^{+69}_{-71}$ Mpc (about four times larger than the GW170817). Another possible reason is that the total mass is so large that the merger promptly collapsed without producing much ejecta. If so, the absence of the HMNS stage (which happens if the mass is less than $\sim 1.5 M_{\rm TOV}$) implies $M_{\rm TOV} \lesssim 2.27 M_\odot \simeq 3.4 M_\odot /1.5 $. This estimate is consistent with the HMNS interpretation of GW170817.

There is also a candidate of BH-NS merger, GW190814 \cite{Abbott:2020khf}, in which the mass for the heavier object is 22.2-24.3 $M_\odot$ and should be a BH, while for the lighter one $2.50$-$2.67M_\odot$. The second one can be either a BH or (rapidly rotating) NS.
If the object is a BH, 
this is a low mass BH in addition to the previously found candidate with the mass $3.3^{+2.8}_{-0.7} M_\odot$ \cite{Thompson:2018ycv}.
If the object is a rotating NS, the EoS must be stiff enough to make the $\simeq 2.6M_\odot$ NS stable; this implies $M_{\rm TOV} \gtrsim 2.08$-$2.23M_\odot \simeq 2.50$-$2.67 M_\odot /1.2$ \cite{Most:2020bba}.

These findings are already quite remarkable, but this is just the beginning. Now we have three powerful GW detectors LIGO-Virgo-KAGRA (KAGRA has just started in 2019), and the IndIGO will start in 2023. There is also the project to build the third generation of detectors, Einstein-Telescope, for 2030s \cite{Maggiore:2019uih}. Finding more events will eventually lead to the statistical analyses, and some events may allow clear-cut interpretations. Also, if a supernova (SN) event happens in our galaxy as in SN1987A, it is possible to obtain GWs \cite{Abdikamalov:2020jzn} together with significant amount of neutrinos; in Hyper-K \cite{Abe:2018uyc}, the expected count of neutrinos is $\sim 10^4$ times greater than for SN1987A. The galactic SN happens a few times in a century, so it might happen within next 10 years.
\\
\\
{\it Neutron Star Interior Composition ExploreR  (NICER) ---} The NICER, launched in 2017, have taken data of X-rays from various pulsars to measure the NS radii and masses simultaneously. The pre-NICER analyses (for a review, see \cite{Watts:2016uzu}) contained rather difficult estimates of the distance to NS, the assumption of black-body radiation, and atmospheric compositions, which introduced the systematic errors. The NICER got rid of the black-body assumption and distance estimate, by following the time evolution of hotspots on NS surface and performing phase-resolved spectroscopy \cite{Watts:2016uzu}. 
Two teams in the NICER individually estimated the radius of the pulsar PSR J0030+0451, reporting $R_1=13.02^{+1.24}_{-1.06}$ km and $M_1 =1.44^{+0.15}_{-0.14}M_\odot $ \cite{Miller:2019cac}, and $R_2=12.71_{-1.19}^{+1.14}$ km for $M_2 = 1.34_{-0.16}^{+0.15} M_\odot$ \cite{Riley:2019yda}. Both estimates tend to be a bit larger than those from GW170817.

In addition, the NICER plans to measure the $2M_\odot$ pulsars, PSR J1614-2230 with the mass $ \simeq 1.91 M_\odot $ \cite{Arzoumanian:2017puf} and PSR J0740+6620 with $\simeq 2.14 M_\odot$ \cite{Cromartie:2019kug}. These measurements will constrain the radii in the high mass region, and hence constrain high density part of EoS.

\section{Delineating dense matter}
\label{sec:eos}

In order to delineate the properties of NS matter, we need studies based on physical pictures on the effective d.o.f. This will remain true even after the $M$-$R$ relations are precisely determined or the sign problem for lattice simulations is solved. For instance, in case of hot QCD, the transition from a hadronic to a QGP was shown to be crossover by lattice simulations \cite{Aoki:2006we}, but the detailed physical picture has emerged after the HRG \cite{Karsch:2003vd} and pQCD \cite{Andersen:2011sf}, with clear-cut effective d.o.f, were used as baselines to diagnose the lattice data \cite{Andronic:2017pug}. It is also desirable to have such baselines to characterize the NS data. In this section we divide the density domain into four categories, based on plausible effective d.o.f. \cite{Kojo:2015fua}.
\\
\\
{\it For $n_B \lesssim 2n_0$ } --- At low density quarks form baryons, whose interactions are mediated by quark-exchanges in the color-singlet channel (meson exchange). Around $\sim n_0$ the strange baryons are absent, and we consider only nucleons (the possibility of very early onset of deconfined quark matter will be mentioned in Sec.\ref{sec:cs2}). The matter is dilute enough for nucleons to exchange only few mesons. 
There are several methods to compute EoS. Two popular approaches seem quite popular: (i) relativistic mean-field (RMF) calculations with in-medium effective interactions; and (ii) non-relativistic many-body calculations based on microscopic {\it bare} nuclear forces.

The RMF approach uses relativistic Lagrangian written in terms of baryon and meson fields \cite{Serot:1984ey}. The (in-medium) coupling constants are arranged to reproduce the nuclear physics near the saturation and finite nuclei.
The clear advantage is its simplicity and flexibility; the method allows us flexible analyses of new experimental data. (For EoS reproducing wide range of phenomenology, see e.g. \cite{Typel:1999yq,Oertel:2016bki,Steiner:2012rk}). Another utility is that the theory includes relativistic effects by construction, and can be used to describe changes in d.o.f. through mean-field variations.  Meanwhile the disadvantage lies in its {\it systematics}; the form of Lagrangian is not well-constrained, as there is no good reason to truncate higher orders of fields due to the lack of counting schemes. The EoS can be sensitive to the choice of higher order couplings.

The approach (ii) first prepares bare two- and three-body forces to reproduce experiments for few-body systems, and then use them for many-body calculations which should be non-perturbative to properly handle soft nucleonic excitations from the nucleon Fermi surface  (however see also Ref.\cite{Kaiser:2001jx} for perturbative approaches). The advantage is its {\it systematics}; there are counting schemes based on either ranges (in potential models \cite{Akmal:1998cf,Togashi:2017mjp}) or momenta of interactions (Chiral Effective Field Theory \cite{Drischler:2017wtt,Lonardoni:2019ypg}). Uncertainties in determining the forces can be directly converted into the EoS, and this sharpens questions on many-body effects. Two-body nuclear forces below the energy $\simeq 350$ MeV are well-constrained by two nucleon scattering experiments, while there are more uncertainties in three nucleon interactions, especially at short distance. In this respect the pure neutron matter computations have less uncertainties because the Pauli blocking does not allow three neutrons to overlap. It is challenging to include other d.o.f. toward higher density, the possible changes in nucleon properties, and relativistic effects which are not small already at $\sim 2n_0$ \cite{Akmal:1998cf}.
\\
\\
{\it For $2n_0 \lesssim n_B \lesssim 5n_0$}  --- Around $2n_0$, many-body forces become increasingly important, casting a doubt on the validity of nuclear matter descriptions. \
Since nuclear forces are mediated by quark exchanges, the importance of many-body forces may lead to the structural changes in nucleons as well \cite{Ma:2020hno}. Meanwhile, the density $\lesssim 5n_0$ is presumably not high enough for establishing quark matter. The domain $2n_0 \lesssim n_B \lesssim 5n_0$ is not well-understood theoretically even at conceptual level, as the effective d.o.f. is not clear-cut. 
There are only a few works  \cite{Lottini:2011zp,Ma:2020hno,Fukushima:2020cmk} that explicitly address the physics in this subtle domain. A recent work \cite{Fukushima:2020cmk} discusses how quark wavefunctions can delocalize through quark exchanges among baryons. 
Assuming that typical quark exchanges take place when meson clouds ({\it valence} quark-antiquark pairs) of the thickness $\sim 0.7$ fm overlap, some modes begin to percolate already around $n_B\sim 1.8 n_0$ (called {\it soft-deconfinement} \cite{Fukushima:2020cmk}), and at higher density more modes are gradually delocalized. 

For practical construction of EoS, there are at least three descriptions. The first is to extrapolate hadronic EoS to high density. Considering the finite size of hadrons this description would be problematic.
 The other two descriptions allow the appearance of quarks. The first is to use nucleonic EoS to $2$-$3n_0$, and switch to quark matter EoS through the first order phase transition, or through the mixed phases in which the transition accompanies various clusters \cite{Glendenning:1992vb,Maslov:2018ghi,Xia:2020brt}. The second possibility is the {\it quark-hadron continuity}  in which hadronic matter is smoothly connected to quark matter. More on this picture will be given in Sec.\ref{sec:cs2}. 
\\
\\
{\it For $5n_0 \lesssim n_B \lesssim 40n_0$}  --- In this regime baryons overlap and the effective d.o.f. should be quarks and gluons, but the density is not high enough for pQCD descriptions. 
This domain has not been studied in detail. 
The density $5n_0 \simeq 0.8\,{\rm fm}^{-3}$ is close to those inside of a single hadron, and hence it is reasonable to expect the validity of quasi-particle descriptions \cite{Leonhardt:2019fua,Song:2019qoh} as in constituent quark models for hadrons. Such constituent quark based discussions \cite{Oka:1980ax,Park:2019bsz} can also explain the baryon-baryon interactions at short distance measured in the lattice simulations \cite{Aoki:2020bew}. Furthermore, the energy density vs {\it mechanical} pressure inside of a proton \cite{Fukushima:2020cmk,Liuti:2018ccr} are found reasonably consistent with neutron star EoS. The mechanical pressure in a proton can be extracted from the gravitational form factor, and can be studied through deeply virtual Compton scatterings \cite{Ji:1996nm}. 
The precise determination of the gravitational form factor is one of important targets in the electron ion collider.

Another hint may be obtained from studies in two-color QCD; for this system lattice MonteCarlo simulations are doable \cite{Boz:2019enj,Iida:2019rah,Astrakhantsev:2020tdl}. There have been seminal works on the phase diagram, EoS, and order parameters. The obvious difference from the three-color case is that baryons are diquarks with which the baryonic matter starts with the Bose Einstein condensation. But such difference may not be essential when density is high; eventually quarks inside of diquarks manifestly establish the Fermi sea, and the condensation remains only around the edge of the Fermi sea as in the BCS theory. Indeed this behavior has been observed on the lattice \cite{Boz:2019enj}. The critical temperature of the diquark condensation is high, $T_s \simeq 100$ MeV, even at a quark chemical potential of $\sim 1$ GeV or $n_B\sim 50n_0$. If we accept the BCS relation $T_s \simeq 0.57 \Delta$ with $\Delta$ being the size of the gap, then $\Delta \simeq 180$ MeV. This would suggest that gluons remain non-perturbative, and the studies of gluon propagators seem consistent with this point of view \cite{Boz:2018crd,Suenaga:2019jjv}.
\\
\\
{\it For $n_B \gtrsim 40n_0$} --- In this domain the pQCD should be valid. The great advantage of this framework is that it contains the error-estimator; by varying the order of $\alpha_s$ and/or renormalization scales, one can infer the importance of higher order corrections. The pioneering calculations were done already in 70's \cite{Freedman:1976ub}, and more systematic error estimates have been carried out in Ref.\cite{Kurkela:2009gj}. The result shows that $\alpha_s$ expansion does not converge well at $\lesssim 40n_0$, indicating that the matter should be strongly correlated. This pQCD EoS have been used as boundary conditions to construct general curves connecting pQCD and nuclear EoS. 

\begin{figure*}[htb]
\vspace{-1.5cm}
\centering
\includegraphics[scale=0.5]{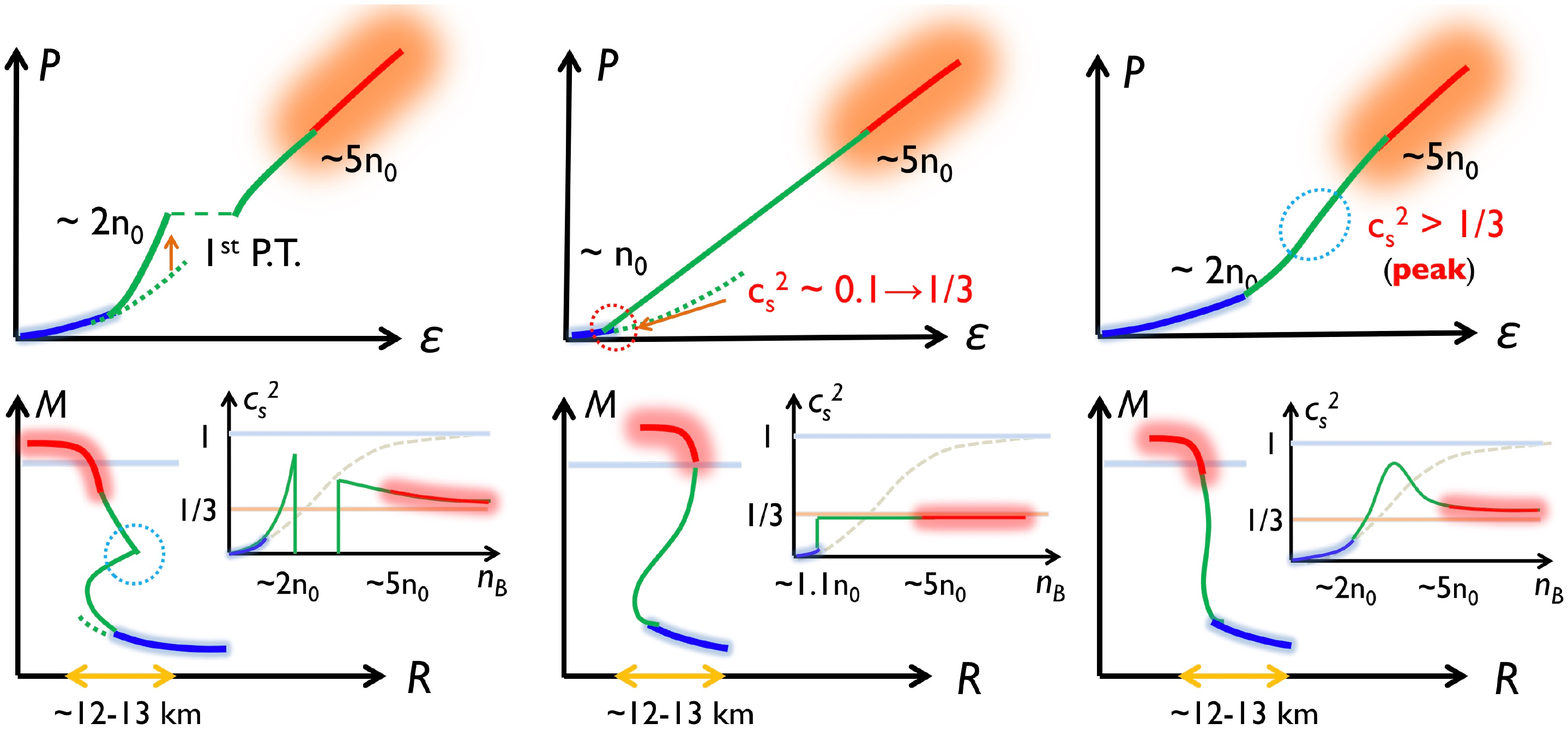}
\vspace{-1.4cm}
\caption{ Rough sketches of $P$-$\varepsilon$, $M$-$R$, and $c_s^2$-$n_B$ relations for three characteristic EoS, see main text for details. For $c_s^2$, we also show the typical behavior of nucleonic EoS with dashed lines.
  }
  \vspace{-0.0cm}
   \label{fig:cs2}
\end{figure*}

\section{Speed of sound}
\label{sec:cs2}

The current major constraints concern with $M_{\rm TOV}$ and $R_{1.4}$.
While they constrain the EoS at different densities, these high and low density parts constrain each other through the causality constraint and thermodynamic stability; these conditions can be summarized into $0\le c_s^2 \le 1$ where $c_s^2= \partial P/\partial \varepsilon$ is the squared speed of sound. The causality constraint becomes severer in the case with a smaller NS radius and a larger $M_{\rm TOV}$, see Refs. \cite{Drischler:2020fvz,Bauswein:2020aag} for comprehensive analyses. Based on the estimates from GW170817 and NICER, below we assume that $R_{1.4} \simeq 12$-$13$ km, excluding very stiff EoS for 1-3$ n_0$. This puts the upperbound on the strength of possible first order phase transition from hadronic to quark matter; if the first order phase transition is too strong, the EoS just after the phase transition are soft, and with the presence of $c_s^2 \le 1$ the pressure cannot grow rapidly enough to satisfy the required stiffness at high density.

This situation differentiates the hadronic-quark transition at high density from the hadron-QGP transition at high temperature. In the latter the $c_s^2$ never exceeds $1/3$ and has a dip in the crossover region; near the transition, many kinds of non-relativistic resonances, with the masses much larger than temperature, can appear because the entropic effects overcome the Boltzmann suppression. A non-relativistic gas naturally has small $c_s^2$ than a pion gas, and this tendency continues until those resonances merge into a QGP, recovering the conformal behavior. In contrast, cold dense matter does not enjoy such entropic effects, and $c_s^2$ can behave quite differently. Since $c_s^2$ in dense QCD may be quite unique, EoS are often parameterized by $c_s^2$ \cite{Alford:2013aca,Miao:2020yjk}.

Shown in Fig.\ref{fig:cs2} are (a bit exaggerated) sketches of $P$-$\varepsilon$, $M$-$R$, and $c_s^2$-$n_B$ relations for three characteristic types of EoS. (In $c_s^2$-$n_B$ relations we also indicate typical nucleonic curves by dashed lines.) The leftmost panel shows the hybrid EoS with the first order phase transition. This case has been most intensively studied. Provided that nuclear EoS are trustable to $n_B \sim 1.5 n_0$, the stiffness must grow so fast that EoS can remain stiff even after the first order phase transition \cite{Benic:2014jia}. Starting with $c_s^2\sim 0.1$ in the nuclear domain, its value exceeds the conformal limit 1/3. The presence of the phase transition is reflected in a kink in the $M$-$R$ curve. In a radical case it may lead to the third family of stars \cite{Alvarez-Castillo:2016oln,Alford:2017qgh,Maslov:2018ghi,Montana:2018bkb}.

The middle case of Fig.\ref{fig:cs2} includes a very rapid growth in $c_s^2$, from $c_s^2 \sim 0.1$ to $\simeq 1/3$ at low density. Assuming quick stiffening to happen at $1.1n_0 \lesssim n_B\lesssim 1.5n_0$, the resulting EoS begin to get stiffened at low density, achieving the required stiffness at high density without invoking $c_s^2 $ greater than $1/3$. It has been known \cite{Freedman:1977gz,Witten:1984rs} that quark EoS with $c_s^2=1/3$ can lead to stiff EoS, if we choose a suitable location for the onset (and neglect the interplay with nuclear matter). The behavior of $c_s^2\simeq 1/3$ is quite different from pure hadronic description, leading to a recent proposal that differentiates quark and hadronic matter by the index $\gamma = c_s^2 \varepsilon/P  $; quark matter is characterized by $\gamma \lesssim 1.75$ \cite{Annala:2019puf}. Substituting $c_s^2 = 1/3$, the condition becomes $P/\varepsilon \gtrsim 0.19$; in the domain $1$-$2n_0$ this ratio is substantially larger than the hadronic case with the pressure suppressed by the nucleon mass. The size of quark matter core is naturally large.

In the rightmost of Fig.\ref{fig:cs2}, nuclear descriptions are trusted to $\sim 2n_0$, so $c_s^2$ does not increase as drastically as in the leftmost and middle panels; with soft EoS at low density, the $2M_\odot$ constraint disfavors first order phase transitions and favors $c_s^2 \gtrsim 1/3$ in some domain; there must be a domain in which $c_s^2$ has a peak that relaxes to the conformal limit, $1/3$, at very large density  \cite{Bedaque:2014sqa,Masuda:2012kf,Kojo:2014rca}. With this picture and the finite size of baryons, the smooth connection of 2-5$n_0$ domain leads to the picture of quark-hadron continuity \cite{Masuda:2012kf,Kojo:2014rca,Ma:2019ery} (whose original proposal comes from the indistinguishability of the hadronic superfluid matter and color-superconducting (CSC) quark matter based on the order parameters \cite{Schafer:1998ef}). Some crossover EoS have been constructed by interpolating hadronic EoS with gapless quark matter \cite{Masuda:2012kf} or CSC quark matter \cite{Baym:2019iky}.

The above three cases remain all compatible with the current observations. What is common for all these cases is the rapid growth in $c_s^2$. It is challenging to construct a theory which leads to such a growing behavior; we note that rapid {\it reduction} of $c_s^2$ can be easily described by the second order-like phase transition, but the rapid {\it enhancement} cannot; this can be seen from the expression
\beq
c_s^2 = \frac{\rmd P}{\rmd \varepsilon} = \frac{ n_B \rmd \mu_B }{\mu_B \rmd n_B} = \frac{n_B}{\mu_B} \bigg( \frac{ \rmd^2 P}{\rmd \mu_B^2 } \bigg)^{-1} \,.
\eeq
With $n_B$ and baryon chemical potential $\mu_B$ being continuous at a transition point, the phase with the bigger susceptibility (or smaller $c_s^2$) is favored. Thus we need mechanisms other than phase transitions with discontinuity. 

One way to cause rapid stiffening is to phenomenologically assume the rapidly growing strong repulsion among baryons. More microscopic arguments were proposed in \cite{McLerran:2018hbz} based on the quarkyonic matter hypothesis \cite{McLerran:2007qj} in which high density matter has the quark Fermi sea but baryonic Fermi surface. Baryons on top of the Fermi sea may be regarded as relativistic baryons with three-quarks collectively moving in the same direction, unlike baryons at low density where the moving directions of three quarks are opposite one another, leaving a small baryon momentum (and pressure) but a large mass density \cite{Kojo:2019raj}. Changes from the non-relativistic to relativistic regime may be a microscopic origin of phenomenological repulsion used in hadronic models. The discussion was initially given for two-flavor matter, but recent extension included the strangeness, with the concept of short-range correlation to make hyperons relativistic \cite{Duarte:2020kvi}.

\section{Summary}
\label{sec:summary}

The progress in NS observations is rapidly improving the constraints on the QCD matter. But even more exciting discoveries would come in next 10 years, thanks to on-going and forthcoming observational programs. Systematic studies are being performed in numerical simulations for NS mergers and SNs.

The $2M_\odot$ constraint, together with low density constraints from nuclear physics and NS radii, suggests that the core density of NS should reach $\gtrsim 5n_0$. Considering the radii of baryons of 0.5-0.8 fm, the dynamics at quark level should be discussed near the NS core.

The natures of hadron-quark phase transitions have not been determined. But the strength of possible first order phase transitions is being constrained. If we find a heavier and more compact NS, the constraint becomes tighter. On the other hand, if we establish kinks in $M$-$R$ curves and associated phenomena, it would readily confirm the existence of the first order phase transition.

The physics of hadron-quark transitions in dense QCD, which are supposed to occur around 2-5$n_0$, are difficult to analyze theoretically because of the fuzzy d.o.f., but a possible strategy in {\it near} future is to improve the constraints at 1-2$n_0$ and 5-40$n_0$. The latter domain would not directly show up in NS phenomenology, but still it constrains EoS in the NS domain.

The speed of sound in dense QCD is likely to be very different from conventional matters. Whatever natures of hadron-quark phase transitions, the $c_s^2$ must either exceed the conformal value 1/3 or increase very rapidly at density 1-2$n_0$. This tendency is certainly different from the hadron-QGP crossover phase transition, and from usual non-relativistic condensed matter with $c_s^2 \ll 1$.

In this article we could not touch thermal effects and general lepton fraction around the core region. Their impacts on the mechanical properties of NS cores may be limited (see however, Ref.\cite{Most:2018eaw} for large latent heat), but they certainly affect the composition around the cores. They can be imprinted, e.g., in neutrino emissions \cite{Nakazato:2010qy,Fischer:2010wp,Fischer:2017lag}. The composition and temperature effects are sensitive to the phase structures, and provide us with very useful tools to diagnose the structure of matter. These physics also have relations to the physics being explored by the low-energy heavy ion collisions which are also expected to have dramatic progress in 2020s due to the activation of new experimental programs \cite{Fukushima:2020yzx,Dexheimer:2020zzs}.

 Many tables and manuals can be found, e.g., in {\it CompOSE} [https\://compose.obspm.fr/home].

\section*{Acknowledgments}
This work was supported by NSFC grant 11650110435.

\bibliographystyle{apsrev4-2}
\bibliography{Ref_AAAPS}

\end{document}